\documentclass[twocolumn]{aastex631}
\usepackage{multirow}
\usepackage{graphicx}
\usepackage{hyperref}
\usepackage{float}

\shorttitle{Raman \ion{O}{6} features in RR Tel}
\shortauthors{Heo et al.}
\graphicspath{{./}{figures/}}
\begin{document}

\title{Raman-scattered \ion{O}{6} Features in the Symbiotic Nova RR~Telescopii \footnote{This paper includes data gathered with the 6.5 meter Magellan Telescopes located at Las Campanas Observatory, Chile.}}

\correspondingauthor{Jeong-Eun Heo}
\email{jeong-eun.heo@noirlab.edu}

\author{Jeong-Eun Heo}
\affil{Gemini Observatory/NSF’s National Optical-Infrared Astronomy Research Laboratory, Casilla 603, La Serena, Chile}
\affil{Departamento de Astronom\'ia, Universidad de La Serena, Av. J. Cisternas 1200 Norte, La Serena, Chile}
\affil{Department of Physics and Astronomy, Sejong University, Seoul, Korea}

 \author{Hee-Won Lee}
\affil{Department of Physics and Astronomy, Sejong University, Seoul, Korea}
 
\author{Rodolfo Angeloni}
\affil{Instituto de Investigaci\'on Multidisciplinar en Ciencia y Tecnolog\'ia, Universidad de La Serena, Av. R. Bitr\'an 1305, La Serena, Chile}
\affil{Departamento de Astronom\'ia, Universidad de La Serena, Av. J. Cisternas 1200 Norte, La Serena, Chile}

\author{Tali Palma}
\affil{Observatorio Astron\'omico, Universidad Nacional de C\'ordoba, C\'ordoba, Argentina}

\author{Francesco Di Mille}
\affil{Las Campanas Observatory, Carnegie Observatories, Casilla 601, La Serena, Chile}



\begin{abstract}

RR~Tel is an interacting binary system in which a hot white dwarf (WD) accretes matter from a Mira-type variable star via gravitational capture of its stellar wind. This symbiotic nova shows intense Raman-scattered \ion{O}{6} 1032 and 1038 features at 6825~\AA\ and 7082~\AA. We present high-resolution optical spectra of RR~Tel taken in 2016 and 2017 with the Magellan Inamori Kyocera Echelle (MIKE) spectrograph at Magellan-Clay telescope, Chile. We aim to study the stellar wind accretion in RR~Tel from the profile analysis of Raman \ion{O}{6} features. 
\replaced{We perform Monte Carlo simulations to reproduce the Raman \ion{O}{6} features by assuming that the \ion{O}{6} emission traces the accretion flow around the WD with a representative scale of $< 0.8~\rm au$.} {With an asymmetric \ion{O}{6} disk model, we derive a representative Keplerian speed of $> 35~{\rm km~s^{-1}}$, and the corresponding scale $< 0.8~\rm au$.}
The best-fit for the Raman profiles is obtained with \deleted{an asymmetric matter distribution of the \ion{O}{6} accretion flow} a mass loss rate of the Mira ${\dot M}\sim2\times10^{-6}~{\rm M_{\odot}~yr^{-1}}$ and a wind terminal velocity $v_{\infty}\sim 20~{\rm km~s^{-1}}$. We compare the MIKE data with \replaced{a previous}{an archival} spectrum taken in 2003 with the Fibre-fed Extended Range Optical Spectrograph (FEROS) at the MPG/ESO 2.2~m telescope. It allows us to highlight the profile variation of the Raman O VI features, indicative of a change in the density distribution \added{of the \ion{O}{6} disk} in the last two decades. We also report the detection of \ion{O}{6} recombination lines at 3811~\AA\ and 3834~\AA, which are blended with other emission lines. \replaced{Based on our profile decomposition, we derive their profile width of $\sim 44~{\rm km~s^{-1}}$. This result implies}{Our profile decomposition suggests} that the recombination of \ion{O}{7} takes place nearer to the WD than the \ion{O}{6} 1032 and 1038 emission region.

\end{abstract}



\section{Introduction} \label{sec:intro}

Symbiotic stars are interacting binary systems consisting of an evolved giant transferring mass to a hot compact object, in most cases a white dwarf (WD) \citep[e.g.,][]{kenyon86}. In the general case, \deleted{of the symbiotic stars,} it is believed that the accretion occurs via the stellar wind from the \replaced{giant rather than Roche Lobe overflow  \citep[e.g.,][]{hillman21}.}{primary, known as Bondi-Hoyle-Lyttleton (BHL) wind accretion.} \added{For the D-type (dust) symbiotic stars, the widest binary systems composing of a Mira-type variable, several hydrodynamical simulations suggest a new mass-transfer model called wind Roche-lobe overflow (WRLOF; \cite{mohamed12}) or gravitational focusing \citep{vkm09}. Accretion through WRLOF focuses on the binary plane, leading to higher mass transfer rate than that of standard BHL wind accretion. This lends support to the suggestion that D-type symbiotic stars are a promising candidate of a type Ia supernova progenitors \citep[e.g.][]{ilk19}.}

RR~Telescopii is a Dusty-type symbiotic nova comprising a Mira component and a WD. A nova-like outburst occurred in 1944 when RR~Tel brightened by $\sim$7~mag in the visual band \citep{ma49}. In 1949, the extended atmosphere started to shrink and entered a nebular phase. Subsequently, its luminosity has been slowly fading for several decades \citep[e.g.,][]{murset91, nuss97}. Recent light curve from the ASAS-SN photometric database shows that the overall V magnitude is still in the steady decline trend. It changed from 11.7 to 11.9 during the last four years \citep{shap14, koch17}. Meanwhile, the effective temperature of the WD has continued to rise from $T \sim 6750~{\rm K}$ in May 1949 \citep{po60} and it has exceeded $> 140000~{\rm K}$ after 1978 \citep[e.g.,][]{hayes96, jor94}. Based on the recent X-ray observations with {\it XMM-Newton}, \cite{gon13} estimated \replaced{the}{its} temperature $T \sim 154000~{\rm K}$ and \deleted{the} luminosity $L\sim5000~{\rm L_{\odot}}$ assuming a distance of $2.6~{\rm kpc}$. The orbital period of this object is still quite uncertain but it is believed to exceed several decades \citep{lee99,sch02}.

From time-series spectroscopic observations of RR~Tel between 1950 and 1960, \cite{tha74} reported a strong, unidentified emission band at 6825~\AA\ having a broad width of $\sim30$~\AA\ with a double-peaked profile. They noted that whenever the 6825 band appeared, the rest of the spectrum showed lines of high ionization, which implied that the origin of the band is related to high-ionized species. A second, weaker band at 7082~\AA\ was also detected,\deleted{ in RR~Tel spectrum,} whose profile bears a striking similarity to 6825 feature \citep{allen79}. Because of their observational properties, including abnormally broad width and the double-peaked profile, \cite{allen79} proposed that both emissions come from the same atomic or molecular species.

\begin{figure}
\centering
\noindent\resizebox{1.2\columnwidth}{!}{
\includegraphics{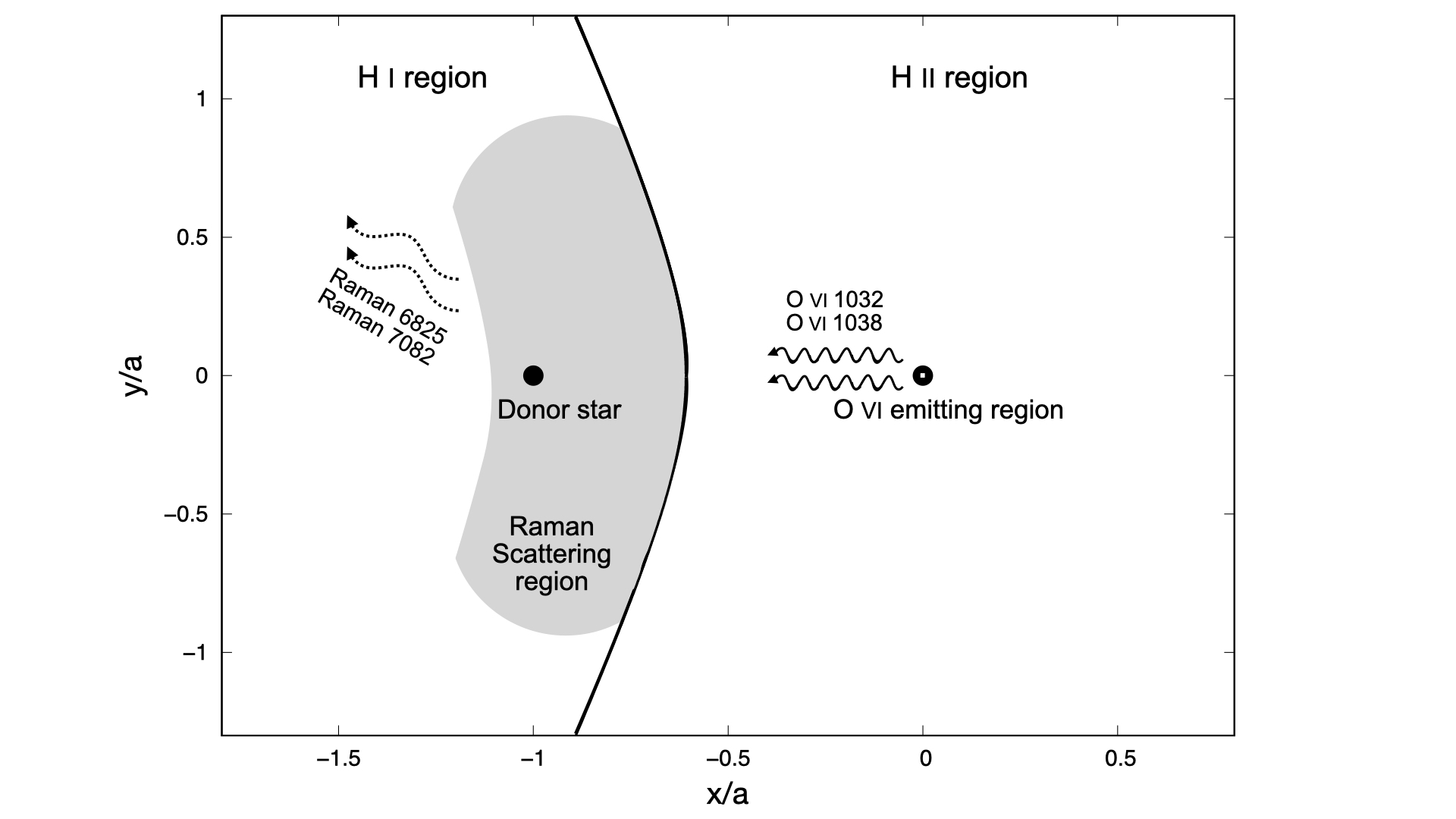}}
\caption{Raman scattering process in symbiotic star. The coexistence of the neutral region characterized by large \ion{H}{1} density in the stellar wind of the donor star(the left black circle) and the strong far-UV source(the right black circle) photoionized by the WD constitutes an optimal condition for Raman scattering by \ion{H}{1}. The black solid line represents the ionization front, determined by a STB geometry. The grey shaded region corresponds the Raman scattering zone in the \ion{H}{1} region.
}
\label{fig:model}
\end{figure}

The identification of the two bands was made by \cite{sc89}, who proposed that the two broad bands at 6825~\AA\ and 7082~\AA\ are due to the Raman scattering of \ion{O}{6}~$\lambda\lambda$ 1032 and 1038 doublet by atomic hydrogen \ion{H}{1}. \replaced{Contemporaneous}{Synchronous} far-UV and optical observations for symbiotic stars including RR~Tel provided support to his identification of the Raman \ion{O}{6} features \citep{espey95,birriel00}. In this scattering process, an \ion{O}{6} 1032 photon is incident on \ion{H}{1} in the ground state, and de-excitation into 2s state produces an optical Raman-scattered photon with $\lambda=$6825~\AA. A similar process with an \ion{O}{6} 1038 photon results in an optical photon at 7082~\AA. Figure~\ref{fig:model} shows the scattering process in symbiotic star. The heterogeneous environment in symbiotic binaries includes the nebular region photoionized by intense far-UV radiation from the hot source and the dense neutral region around the donor star, providing an excellent laboratory to study Raman scattering by \ion{H}{1} \citep[e.g.,][]{sc89,nu89}.

The inelasticity of Raman scattering requires that the wavelength of an emergent Raman photon is almost solely determined by the relative motion between the far-UV \ion{O}{6} emission region and the \ion{H}{1} scattering region and negligibly dependent of the observer's line of sight. This unique property allows us to investigate the mass transfer process with an edge-on view of the cool star. Follow-up spectropolarimetric observations revealed that the Raman \ion{O}{6} bands of RR~Tel are strongly polarized and show complicated line structures with a polarization flip in the red wing \citep{sc94,harries96,sc99}. The spectropolarimetric properties were explained by a very extended scattering region responsible for a large mass loss rate. Another model was suggested by \cite{lee99}, who adopted an accretion disk around a WD with an asymmetric matter distribution, resulting in the asymmetric double-peak profiles of the Raman \ion{O}{6} in RR~Tel. Previous research works on line formation of Raman \ion{O}{6} features in other symbiotic stars show that their profiles provide valuable kinematic information on the accretion \replaced{flow}{disk}. In particular, the main peak separation of Raman \ion{O}{6} features indicates a representative velocity scale $\sim50~\rm {km~s^{-1}}$ of the accretion \replaced{flow}{disk} around the WD \citep[e.g.,][]{lee07,heo15,heo16}. 

Further interesting point of Raman \ion{O}{6} is that the two profiles, as usual, are not identical in a given object, even though the \ion{O}{6} $\lambda \lambda$ 1032 and 1038 photons are formed in the same region. The difference in flux between the two peaks is more evident in 7082 feature than 6825 profile. \ion{O}{6} $\lambda \lambda$ 1032 and 1038 resonance doublet lines arise from $2s_{1/2}-2p_{3/2,1/2}$ transitions. Since the oscillator strength of $2s_{1/2}-2p_{3/2}$ transition is twice larger than that of $2s_{1/2}-2p_{1/2}$ transition, it is expected that the flux ratio $F(1032)/F(1038)=2$, in the optically thin region. In accordance with the ratio of oscillator strength, \ion{O}{6} 1032 is twice more optically thick than \ion{O}{6} 1038, $\tau(1032)=2\tau(1038)$, which makes it more difficult for an \ion{O}{6} $\lambda$ 1032 line photon to escape from the site. We, therefore, expect that the flux ratio $F(1032)/F(1038)$ may deviate from the optically thin limit of 2 and approach unity in an optically very thick medium. \cite{heo15} adopted the local variation of the flux ratio $F(1032)/F(1038)$ in the accretion \replaced{flow}{disk} to account for the profile disparity of the Raman \ion{O}{6} features in a symbiotic nova V1016~Cyg. As demonstrated by these results, Raman \ion{O}{6} features can be an ideal spectroscopic tool to \replaced{look into}{investigate} the vicinity of the WD and further probe the mass transfer process associated with stellar wind accretion in symbiotic stars.

Although RR~Tel is a well-studied symbiotic nova, there is a lack of optical studies for the past two decades. The most recent observation in the literature was conducted in 2000, analyzed by \cite{young12}. In this paper, we present and discuss our high-resolution spectroscopic observations of the Raman \ion{O}{6} features in RR~Tel, conducted in 2016 and 2017 using the Magellan Inamori Kyocera Echelle (MIKE) spectrograph installed on the 6.5~m Magellan-Clay telescope, Chile. \added{In Section~\ref{sec:obs},} we investigate the Raman \ion{O}{6} features comparing the archival far-UV data taken in 2002 with the { \it Far-Ultraviolet Spectroscopic Explorer (FUSE)} satellite and the almost contemporary optical spectrum taken in 2003 with the Fibre-fed Extended Range Optical Spectrograph (FEROS) at the MPG/ESO 2.2 m telescope. A detailed profile analysis of the Raman \ion{O}{6} features based on the accretion \replaced{flow}{disk} model was performed, and the results are shown in Section~\ref{sec:sim}. Our finding of the \ion{O}{6}~$\lambda\lambda$ 3811 and 3834 doublets from our profile decomposition is described in Section~\ref{sec:ovi381134}.
Discussion and summary follow in Sections~\ref{sec:dis} and \ref{sec:sum}.

\section{Observations} \label{sec:obs}
\explain{The order of subsections has been changed.}

\subsection{MIKE Spectrum}\label{sec:mike}
We obtained high-resolution optical spectra of RR~Tel using the MIKE spectrograph \citep{bernstein03}. The double echelle spectrograph MIKE covers the wavelength range of 3350-5000~\AA\ (blue) and 4900-9500~\AA\ (red) in its normal configuration. A slit width of $0.7''\times5''$ was used, resulting in resolving power $R \sim 27000$ and $\sim 35500$ on the blue and red sides, respectively. To increase the signal-to-noise ratio (S/N), a $2\times2$ binning was applied in both the spectral and spatial directions. Observations were carried out on 2016 July 30 and 2017 July 26, for total exposure times of 2000~sec and 2400~sec, respectively. With these integration times, the spectra have significant $S/N > 100$ for strong lines suitable for profile analysis. ThAr spectra were taken for wavelength calibration. 

We reduced the MIKE data using the Carnegie Python Distribution (CarPy) pipeline \citep{kelson00,kelson03} and \texttt{NOAO.onedspec} package of the Image Reduction and Analysis Facility (IRAF).\footnote{IRAF is distributed by the National Optical Astronomy Observatory, which is operated by the Association of Universities for Research in Astronomy (AURA) under cooperative agreement with the National Science Foundation.} The data are overscan-corrected, bias-subtracted, extracted, wavelength calibrated, and flat-fielded with quartz lamps within the pipeline. Individual exposures were then combined using the IRAF task \texttt{scombine}. Due to the poor weather condition of the night 2016 July 30, flux calibration was not possible. For the observing night 2017 July 26, we observed the spectrophotometric standard star HR~7950 under photometric condition. Flux calibration was performed with the IRAF tasks \texttt{standard, sensfunc}, and \texttt{calibrate}.

\begin{figure*}
\centering
\includegraphics[scale=0.3]{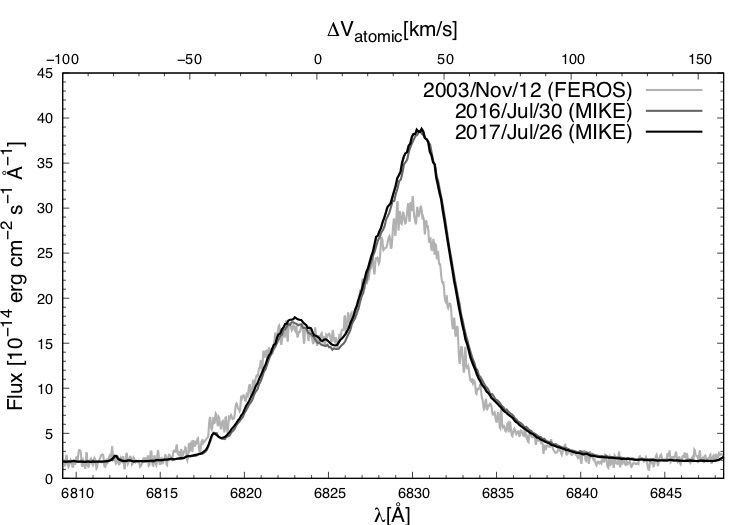}
\includegraphics[scale=0.3]{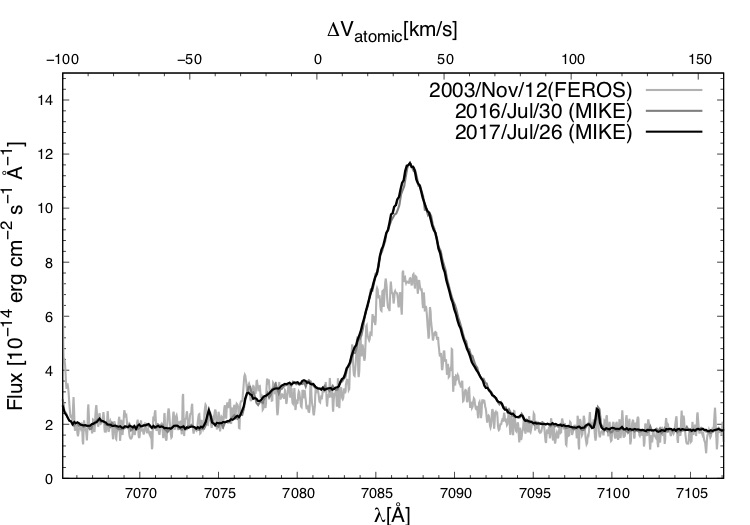}
\caption{The Raman-scattered \ion{O}{6} $\lambda\lambda$ 1032 and 1038 features at 6825 (left) and 7082~\AA\ (right) in the spectra of RR~Tel. Light grey lines represent the FEROS data, and the medium grey and black lines show the 2016 and 2017 MIKE data, respectively. The upper horizontal axis shows the Doppler factor $\Delta V_{atomic}$ computed in the parent far-UV \ion{O}{6} space. The 2003 and 2016 spectra are multiplied by an artificial factor for profile comparison with the flux-calibrated 2017 data.
}
\label{MIKE-FEROS}
\end{figure*}

The left and right panels of Fig.~\ref{MIKE-FEROS} show the Raman-scattered \ion{O}{6} features at 6825~\AA\ and 7082~\AA. To make quantitative profile comparisons of the two Raman \ion{O}{6} features, we transform the observed wavelengths into the \ion{O}{6} Doppler factor $\Delta V_{atomic}$ as shown by the upper horizontal axis. First, we convert the observed wavelength to the vacuum wavelength using the refractive index of air $n_{air}=1.0003$. By measuring the center wavelengths of the strong emission lines (e.g., \ion{He}{1}$~\lambda$7065 and $\lambda$6678, [\ion{N}{2}]~$\lambda$6583 and $\lambda$6548, [\ion{O}{3}]~$\lambda$4959, H$\beta$ and H$\gamma$), we obtain the systemic velocity $\nu_{sys}$ of $-55.5~{\rm {km~s^{-1}}}$, and $-57.8~{\rm {km~s^{-1}}}$, respectively, for the 2016 and 2017 data. We correct for the systemic velocity and calculate the corresponding far-UV wavelength of incident \ion{O}{6} photon using the energy conservation principle. Adopting the \ion{O}{6} center wavelengths of  $\lambda_{1032}=1031.928$~\AA\ and $\lambda_{1038}=1037.618$~\AA, we finally compute the \ion{O}{6} parent Doppler factor $\Delta V_{atomic}$. More detailed information about the transformation between the wavelength and the Doppler factor is available in \cite{heo16}.

We fit the Raman \ion{O}{6} features with two Gaussian components, from which we derive the peak separation of the Raman \ion{O}{6} 6825 feature $\Delta V=47.5~{\rm km\ s^{-1}}$ and $\Delta V=48.7~{\rm km\ s^{-1}}$ in the data of 2016 and 2017, respectively. For the Raman \ion{O}{6} 7082 features, the peak separations are measured to be $\Delta V=45.7~{\rm km\ s^{-1}}$ and $\Delta V=46.4~{\rm km\ s^{-1}}$ in 2016 and 2017, respectively. The Gaussian fitting parameters are presented in Table~\ref{tb:Raman_gauss}.

\begin{figure}
\centering
\includegraphics[scale=0.3]{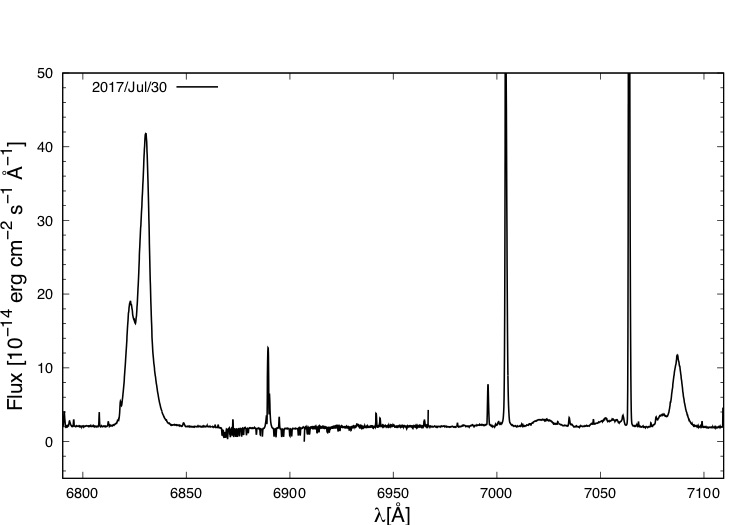}
\caption{Part of MIKE spectrum of RR~Tel around Raman-scattered \ion{O}{6} features at 6825~\AA\ and 7082~\AA\ taken in July 2017. This region has a constant continuum value of $2.0\times~10^{-14}~{\rm erg~cm^{-2}~s^{-1}~\AA^{-1}}$, which is not contaminated by the Mira component. The Raman \ion{O}{6} features are located away from the telluric bands at 6870~\AA.
}
\label{mike2017}
\end{figure}

We overplot the MIKE data in Fig.~\ref{MIKE-FEROS}, whose medium grey lines represent the 2016 data and black lines show the 2017 data.
The vertical axes show flux density in units of $10^{-14}~{\rm erg~cm^{-2}~s^{-1}~\AA^{-1}}$. Since the 2016 data is not flux calibrated, we multiplied an artificial factor to match the continuum value with the 2017 data. As shown in Fig.~\ref{mike2017}, the continuum around the bands is virtually flat with a constant value of $2.0\times~10^{-14}~{\rm erg~cm^{-2}~s^{-1}~\AA^{-1}}$ due to the dust obscuration. The Mira in RR Tel is heavily obscured and so no trace of photospheric absorptions is visible \citep[e.g.][]{kot06}. Dust obscuration in RR~Tel is also supported by its optical light curves, which do not show any pulsation of Mira, whereas the Near-IR light curves show a clear periodicity \citep{groma09}.

To measure the total line fluxes for the Raman \ion{O}{6} features, we assume that the emission having the Doppler factor $\Delta V_{atomic}$ between $-60~{\rm {km~s^{-1}}}$ to $100~{\rm {km~s^{-1}}}$ is a Raman \ion{O}{6} feature. In these regions, the contributions of other spectral lines, including TiO or VO absorption lines from the Mira and telluric lines, are negligible. There is telluric ${\rm O_{2}}$-B band between 6870 - 6950 \AA, but not affecting both Raman \ion{O}{6} features \citep{groppi96}. Few unknown emission lines with FWHM $<0.5$~\AA\ are also detected at the wavelength range of Raman \ion{O}{6} features. However, their contribution to the total line fluxes are negligible due to the relatively small intensities compared with the Raman features. Therefore we obtain the total line fluxes of the Raman \ion{O}{6} features $F(6825)=30.34 \pm 0.11 \rm \times10^{-13}~erg~cm^{-2}~s^{-1}$ and $F(7082)= 5.98 \pm 0.08 \rm \times10^{-13}~erg~cm^{-2}~s^{-1}$ for the 2017 data.

\begin{table}
\caption{Observed line fluxes of the Raman \ion{O}{6} 1032 and 2018 features at 6825 and 7082 \AA\ from 1993 to 2017. \label{tb:raman}}
\vskip5pt
\noindent\resizebox{1.0\columnwidth}{!}{
 \begin{tabular}{cccc}
 \hline \hline
Date & Raman \ion{O}{6} 1032 & Raman \ion{O}{6} 1038 & ref. \\
& ($\rm{erg\ cm^{-2}\ s^{-1}} $)  & ($\rm{erg\ cm^{-2}\ s^{-1}} $)  & \\
\hline
1993-11-21 & $ 7.2\times10^{-12}$ & $1.2\times10^{-12}$ & \cite{sc99} \\
1995-03-18 & $6.7\times10^{-12} $ & $1.3\times10^{-12} $ & \cite{espey95}\footnote{extinction-corrected flux} \\
1996-07-22 & $ 4.78\times10^{-12}$ & $8.98\times10^{-13}$ & \cite{craw99} \\
1996-10-08 & $4.5\times10^{-12}$ & $9.1\times10^{-13}$ & \cite{sc99} \\
2017-07-26 & $3.03 \pm 0.01 \times10^{-12}$ & $5.98 \pm 0.08 \times10^{-13}$  & This work \\
 \hline \hline
  \end{tabular}
  }
\end{table}

The Raman line fluxes of RR~Tel were studied in previous observations between 1993 and 1996  \citep[e.g.,][]{espey95,craw99,sc99}. A steady decrease was found over the period, and the same behavior is shown for the far-UV \ion{O}{6} 1032 and 1038 emissions. We note that the fluxes of both Raman \ion{O}{6} in the MIKE data have decreased by about 30~\% from 1996 October. The measurements are summarized in Table~\ref{tb:raman}. The decrease in the line intensity is consistent with the trend of general slow fading started in 1949 after its outburst \citep[e.g.,][]{nuss97,con99}.

\subsection{FUSE Spectrum} \label{sec:fuse}

RR~Tel was observed on June 14, 2002 with the {\it FUSE} satellite as part of the Cycle 3 program C141 (\added{data ID: C1410102000;} PI: P.R. Young). The {\it FUSE} spectrum of RR~Tel was taken for an exposure time of 128~sec through the low-resolution aperture (LWRS: 30 $\times$ 30 arcsec) using  TTAG (Photon Address) mode. The SiC2 channel covers a wavelength range of 917 - 1104~\AA\ at a spectral resolution of 12000 - 20000 \citep{moos00,sah00}. The data were re-processed and archived at MAST with the CalFUSE pipeline version 3.2.1. The extracted data is binned in wavelength with a bin size of 0.013~\AA\ \citep{dix07}. Considering that the observation was made under low-resolution mode, we binned the data by a spectral resolution of 0.362~\AA.

\begin{figure}
\centering
\includegraphics[scale=0.3]{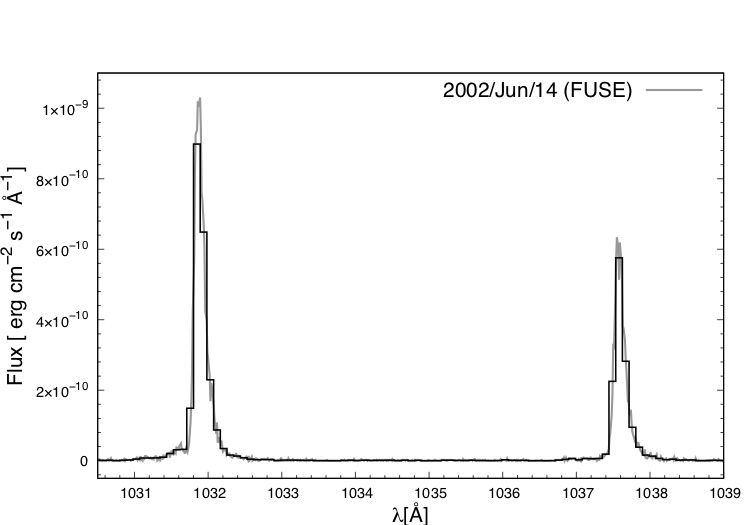}
\caption{Part of the {\it FUSE} spectrum of RR~Tel around \ion{O}{6}~$\lambda\lambda$~1032 and 1038 taken in June 2002. The light grey line represents the raw data, whereas the black line shows the binned spectrum.
}
\label{fig:fuse}
\end{figure}

\begin{table*}
\centering
\caption{Observed line flux of \ion{O}{6}~$\lambda\lambda$~1032 and 1038 resonance doublet and their flux ratio from 1993 to 2002. \label{tb:fuse}}
\vskip5pt
 \begin{tabular}{ccccc}
 \hline \hline
Date & \ion{O}{6} 1032 & \ion{O}{6} 1038 & F(1032)/F(1038) & ref. \\
& ($\rm{erg\ cm^{-2}\ s^{-1}} $)  & ($\rm{erg\ cm^{-2}\ s^{-1}} $)  & & \\
\hline
1993-09-16 & $2.56 \times 10^{-10}$ & $1.54 \times 10^{-10}$ & 1.70  & \cite{sc99} \\
1995-03-12 & $2.32 \times 10^{-10}$ & $1.43 \times 10^{-10}$ & 1.64 & \cite{birriel00} \\
2002-06-14 & $2.01 \pm 0.56 \times 10^{-10}$ & $1.18 \pm 0.32 \times 10^{-10}$ & 1.70 & This work \\
 \hline \hline
  \end{tabular}
\end{table*}

In Fig.~\ref{fig:fuse}, we present a zoom-in of the {\it FUSE} spectrum centered on the two strongest emission lines, \ion{O}{6} 1032 and 1038. The light grey line represents the raw data, whereas the black line shows the binned spectrum. The \ion{O}{6} emissions have line fluxes of $F(1032)=2.01 \pm 0.56 \times10^{-10}~\rm erg\ cm^{-2}\ s^{-1}$ and $F(1038)=1.18 \pm 0.32 \times10^{-10}~\rm erg\ cm^{-2}\ s^{-1}$, respectively, yielding the flux ratio of $F(1032)/F(1038)=1.70$. There has been a steady decrease in line fluxes of both \ion{O}{6} emissions from $F(1032)=2.56\times10^{-10}~\rm erg\ cm^{-2}\ s^{-1}$ and $F(1038)=1.54\times10^{-10}~\rm erg\ cm^{-2}\ s^{-1}$ in 1993 \citep{sc99} - a drop of 22~\%. We summarize the observed values of \ion{O}{6} lines and compare them with previous measurements from other research works in Table.~\ref{tb:fuse}. We note that the flux ratio $F(1032)/F(1038)$ has remained almost constant at $\sim$ 1.7 throughout this period.

\subsection{FEROS Spectrum}

\begin{table*}
\caption{Fitting parameters (center wavelength $\lambda$, corresponding Doppler factor $V$ and peak value $f$) for the Raman \ion{O}{6} features through a two-component Gaussian modeling. \label{tb:Raman_gauss}}
\vskip5pt
\centering
\noindent\resizebox{1\textwidth}{!}{
\begin{tabular}{ccccccccc}
\hline \hline
 & $\lambda_{blue}$ & $V_{blue}$ & $f_{blue}$ & $\lambda_{red}$ &  $V_{red}$ & $f_{red}$ & $\Delta V$ & $f_{Red}/f_{blue}$ \\ 
 & (\AA) & $(km s^{-1})$  & ($\times10^{-13}\rm{erg\ cm^{-2}\ s^{-1}\ \AA^{-1}}$) &  (\AA) &  ($km\ s^{-1}$) & ($\times10^{-13}\rm{erg\ cm^{-2}\ s^{-1}\ \AA^{-1}}$)  & ($km\ s^{-1}$)  & \\ \hline
Raman 6825 & \multicolumn{8}{c}{} \\
2003-11-13 & 6822.75 & -10.0 & $1.40$ & 6829.75 & 36.5 & $2.75$ & 46.5 & 1.96 \\ 
2016-07-30 & 6823.10 & -8.2 & $1.50$ & 6830.25 & 39.3 & $3.65$ & 47.5 & 2.43 \\ 
2017-07-26\footnote{\label{note1} flux-calibrated} & 6822.90 & -9.2 & $1.55$ & 6830.20 & 39.5 & $3.67$ & 48.7 & 2.37 \\ \hline
Raman 7082 & \multicolumn{8}{c}{} \\ 
2003-11-13 & 7978.70 & -15.1 & $0.14$ & 7086.60 & 33.7 & $0.56$ & 48.8 & 4.00 \\ 
2016-07-30  & 7079.90 & -8.3 & $0.17$ & 7087.30 & 37.4 & 0.98 & 45.7 & 5.76 \\
2017-07-26\textsuperscript{\ref{note1}} & 7079.70 & -9.3 & $0.17$ & 7087.20 & 37.1 & $0.98$ & 46.4 & 5.76 \\ \hline \hline
\end{tabular}
}
\end{table*}

RR~Tel was observed with FEROS installed on the MPG/ESO 2.2~m telescope for a total exposure time of 600~sec on November 13, 2003. The spectrum spans the wavelength region 3527 - 9216~\AA\ and has a spectral resolution of $R\sim 48000$ \citep{Kaufer99}. The data were reduced automatically using the FEROS Data Reduction Software (DRS) pipeline version fern/1.0 and archived at the European Southern Observatory (ESO) database. \footnote{\url{http://archive.eso.org/scienceportal/home}} The data are overscan-corrected, bias-subtracted, extracted and flat-fielded. The extracted, flat-fielded data are wavelength calibrated and rebinned in wavelength with a bin size of 0.3~\AA. The reduced spectra are not flux calibrated. \footnote{\url{http://www.eso.org/rm/api/v1/public/releaseDescriptions/94}} In Fig.~\ref{MIKE-FEROS}, the light grey lines correspond to the FEROS data. Not being flux calibrated, the FEROS data is multiplied by an artificial factor to compare their profiles with that of the flux calibrated MIKE data.

Following the same method used for the MIKE data, we measure the center of the strong emission lines (\ion{He}{1}$~\lambda$7065 and $\lambda$6678, [\ion{O}{3}]~$\lambda$5007 and $\lambda$4959), and obtain the systemic velocity $\nu_{sys}$ of $-59.7~\rm {km~s^{-1}}$. Through two-component Gaussian modeling of the Raman features, we find that the two peaks of the Raman \ion{O}{6} 6825 are $\lambda_{blue}=6822.75$~\AA\ and  $\lambda_{red}=6829.75$~\AA. These correspond to a Doppler factor of $V_{blue}=-10.0~{\rm km\ s^{-1}}$ and $V_{red}=36.5~{\rm km\ s^{-1}}$ resulting in the peak separation $\Delta V=46.5~{\rm km\ s^{-1}}$. For the Raman \ion{O}{6} 7082, the peak separation $\Delta V=48.5~{\rm km\ s^{-1}}$ is obtained. We summarize the Gaussian fitting parameters for the Raman features in Table~\ref{tb:Raman_gauss}.

\subsection{Profile Variation of Raman \ion{O}{6}}\label{sec:profile}
In previous studies, it was found that the Raman \ion{O}{6} features of RR~Tel show an asymmetric double-peaked profile, where the red peak is stronger than the blue one \citep[e.g.,][]{allen80,sc99}. \cite{harries96} investigated the temporal variation of Raman \ion{O}{6} 6825 feature between 1992 and 1994, from which they found a slight decrease in (normalized) line intensity while the overall profiles remained unchanged (see Fig.~8-d in \cite{harries96}).

By comparing the data obtained in the years of 2003, 2016, and 2017, we find that both Raman \ion{O}{6} profiles show significant variations. In Fig.~\ref{MIKE-FEROS}, we overplot the FEROS spectrum taken in 2003 and the MIKE data observed in 2016 and 2017. It is obvious that the asymmetry of both Raman profiles has grown for the last two decades, whereas there is no significant change between 2016 and 2017. To describe the profile asymmetry in a given Raman feature, we use parameter $f_{red}/f_{blue}$, the ratio between the peak values of the two Gaussian components. We find that the asymmetry parameter $f_{red}/f_{blue}$ of Raman \ion{O}{6} 6825 increased $26~\%$ from $1.94$ in 2003 to $2.45$ in 2016. This change is even more apparent in the 7082 feature, a $43~\%$ increase from $3.94$ in 2003 to $5.65$ in 2016.

\section{Line formation of Raman \ion{O}{6}} \label{sec:sim}
We introduce our line formation modeling via a Monte Carlo approach to compute the flux and the profiles of Raman-scattered \ion{O}{6} features. 
The Monte Carlo simulation starts with a generation of a far-UV \ion{O}{6} photon at a random place in an accretion \replaced{flow}{disk}. The \ion{O}{6} photon enters the \ion{H}{1} region \replaced{and wanders in the region with Rayleigh scattering.}{where is Rayleigh scattered.} It escapes the region if the photon has an optical depth $\tau$ larger than $\tau_{\infty}$, the scattering optical depth to an observer at infinity, or becomes an optical photon through Raman scattering. 
A full description of the simulations, including the escape condition and the scattering process, can be found in \cite{lee97}.

It is worth \replaced{highlighting}{remembering once again} that the line profiles of Raman-scattered features are almost independent of the observer's line of sight and are almost solely determined by the relative motion between the far-UV \ion{O}{6} emission region and the \ion{H}{1} scattering region. Because the \ion{H}{1} region around the Mira commands an almost perfect edge-on view of the accretion around the WD, the profile analysis of Raman-scattered features provides crucial information, including physical extent, density distribution, and kinematics of the \ion{O}{6} 1032 and 1038 emission region around the WD \citep{lee07,heo16,lee19}.

\subsection{Stellar Wind and the STB Ionization Front}

First of all, to model the density distribution of \ion{H}{1} around the Mira, it is assumed that the stellar wind from the Mira follows a beta law, where the radial wind velocity $v_r({\bf r})$ is given by
\begin{equation}
v_r({\bf r}) = v_\infty (1-R_*/r)^\beta.
\end{equation}
Here, $R_*$ is the launching site of the wind, $v_\infty$ is the wind terminal velocity \citep[e.g.,][]{lamers99}. We choose the parameter $\beta=1$ for simplicity in our Monte Carlo simulations. 

The slow stellar wind around the Mira is illuminated by intense far-UV radiation from the hot component, and hence some parts facing the WD are photoionized. \cite{stb84} presented their photoionization calculation to find the ionization front, known as the STB model. In the STB model, the ionization front in the stellar wind region is determined by the balance of photoionization rate by the H-ionizing luminosity from the hot component and recombination rate set by the mass-loss rate. 

In the STB geometry, the ionization front is specified by the ionization parameter $X$, which is given by
\begin{equation}
X={4\pi a L_{H}\over\alpha_{B}}{\left({m_{H}v_{\infty}\over~\dot{M}}\right)}^2.
\end{equation}
Here, $a$ is the binary separation, $L_{H}$ is the H-ionizing photon number luminosity, $\alpha_{B}$ is the case B recombination coefficient for \ion{H}{1}, and $m_{H}$ is the proton mass. In this work, we adopt the values of $a=56 \rm~au$, $L_{H}=7\times 10^{47}~\rm~s^{-1}$ that were used in previous works \citep{hin13,gon13}.

\subsection{An Asymmetric \ion{O}{6} Accretion Disk}

\added{Many hydrodynamical simulations of stellar wind accretion in detached binaries involving an evolved giant star show that \deleted{the} stable disks can form around WD \citep[e.g.,][]{ma98,vkm09,huarte13}.}
\replaced{Our MIKE data shows that the Raman \ion{O}{6} features have double-peaked profiles, from which we identify that the far-UV \ion{O}{6} 1032 and 1038 emission region is part of the accretion flow around the WD.}{In view of the fact that the Raman \ion{O}{6} profiles are dependent exclusively on the relative motion between the emitter and scatterer, the double-peak profiles of Raman \ion{O}{6} strongly imply that the far-UV \ion{O}{6} emission lines are formed in regions that move toward the \ion{H}{1} region and recede from it. In this work, the far-UV \ion{O}{6} 1032 and 1038 emission regions are identified as a part of the accretion disk formed around the WD.}
\added{Since Raman \ion{O}{6} profiles reflect only relative kinematics between the \ion{O}{6} emission region and the \ion{H}{1} scattering region, but not the radial motion nor 3D distribution, a simple circular disk at $z=0$ is used in our simulation. We also assume that the disk follows a Keplerian velocity profile.
Our assumption is supported by hydrodynamical simulations, which show that the orbital speed in the disk is close to Keplerian \citep[e.g.,][]{vkm09}.}

\added{Wind accretion can lead to an asymmetry of the accretion flow around WD with significant density enhancement of the entering side, associated with flow streamlines, bow shocks, and ellipticity, etc \citep[e.g.,][]{vkm09,huarte13,sal18}.}
We\replaced{also}{, therefore,} attribute the red-peak enhanced \added{Raman \ion{O}{6}} profiles to \added{the} asymmetry of the matter distribution in the \ion{O}{6} \replaced{emission region}{disk} \citep[e.g.,][]{heo15}. 
\deleted{A number of hydrodynamic studies lend support to the formation of an accretion disk around the WD, where the accretion flow is quite asymmetric with significant density enhancement on the entering side \citep[e.g.,][]{ma98,vkm09}.}
 The part of the accretion \replaced{flow}{disk} on the entering side is moving away from the Mira and hence is characterized by positive Doppler factors for Raman-scattered \ion{O}{6}: we call this region the red emission region (RER) and the opposite side the blue emission region (BER).  

It is noteworthy that the profiles of the two Raman \ion{O}{6} features are not identical: \deleted{more} specifically, the ratio of the red and blue peak fluxes $f_{red}/f_{blue}$ is 2.4 for the Raman 6825 whereas it is 5.8 for the Raman 7082 as shown in Section.~\ref{sec:profile}. The disparate profiles of the two Raman \ion{O}{6} features can be explained by the local variation of the ratio $F(1032)/F(1038)$ in the \ion{O}{6} emission region, which decreases from 2 to 1 as optical depth increases \citep{heo15}. The {\it RER}, assumed to be of high density, is characterized by the flux ratio $F (1032)/F (1038)\sim1$, whereas the {\it BER} is much more sparse, resulting in $F(1032)/F (1038)\sim2$.

\subsection{Monte Carlo Results}

\begin{figure*}[ht!]
\centering
\includegraphics[scale=0.14]{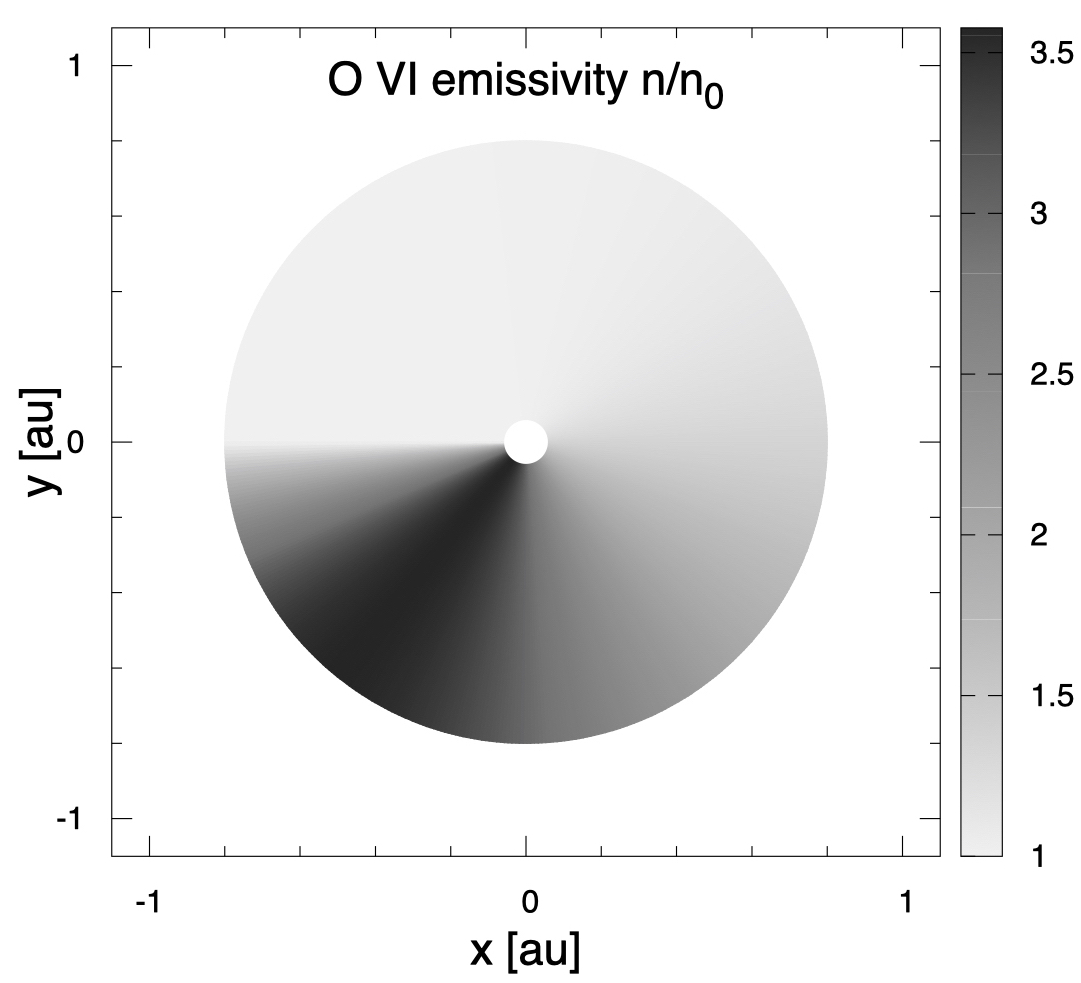}
\includegraphics[scale=0.14]{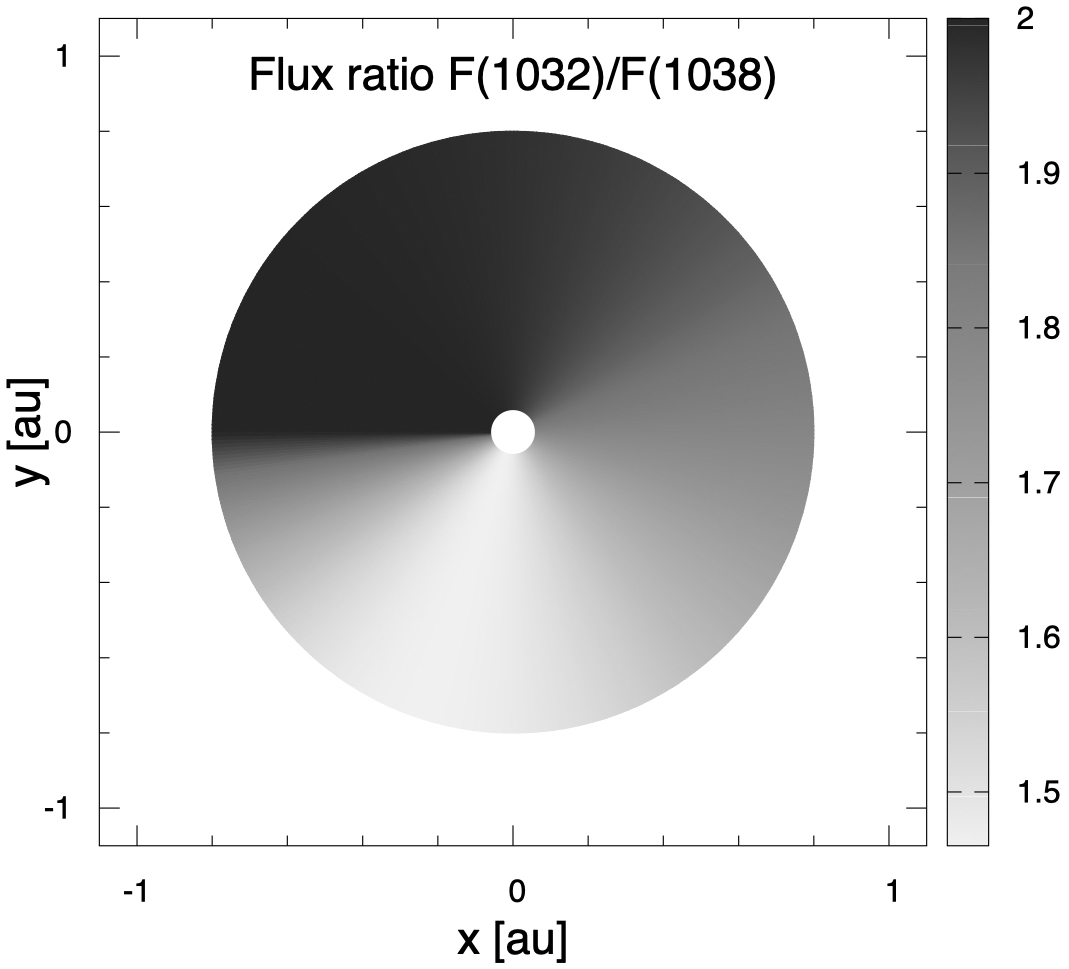}
\includegraphics[scale=0.27]{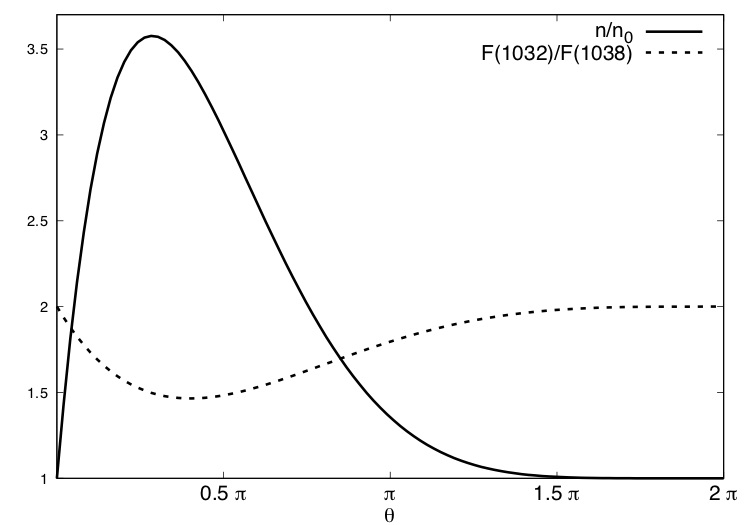}
\caption{
\explain{The figure has been changed}
\added{Schematic top-view of the \ion{O}{6} emitting region around the WD with the \ion{O}{6} emissivity (left) and the flux ratio F(1032)/F(1038) (middle).} The asymmetry of the matter distribution in the \ion{O}{6} region is given as a function of the azimuthal angle $\theta$, where $\theta=0$ coincides with the direction from the WD to the Mira. The corresponding density function $n(\theta)/n_{0}$ (solid line) and the flux ratio $F(1032)/F(1038)$ (dotted line) are shown in the right panel.
}
\label{fig:denf}
\end{figure*}

\deleted{This work assumes the accretion flow to be Keplerian, and azimuthally asymmetric.} To reproduce \replaced{the observed profiles}{the peak separation} of the Raman \ion{O}{6} features, the Keplerian velocity is set to $> 35~{\rm km~s^{-1}}$. It corresponds to the physical size of $<0.8$~au when we adopt $M_{WD}=0.65~{\rm M_{\odot}}$ as the mass of the WD \citep{gon13}. The asymmetry of the matter distribution in the accretion \replaced{flow}{disk} is given as a function of the azimuthal angle $\theta$  measured from the line toward the Mira.
The best-fit is made using the following density function, 
\begin{equation}
  n(\theta)=n_{0}(1+\theta \times {(2\pi-\theta)^6}/8500).
 \end{equation}
\added{In Fig.~\ref{fig:denf}, we present the schematic top-view of the \ion{O}{6} disk with the corresponding \ion{O}{6} emissivity (left).} \replaced{In Fig.~\ref{fig:denf},}{In the right panel,} the solid line represents the \ion{O}{6} density profile $n(\theta)/n_{0}$ as a function of $\theta$. The \ion{O}{6} emissivity is dominant in the {\it RER} ($0<\theta<\pi$), whose peak is at around $\theta=0.3\pi$. 

Considering the disparate profiles of the two Raman \ion{O}{6}, we use the flux ratio $F(1032)/F(1038)$, varying from 1.5 to 2, as a function of $\theta$
 \begin{equation}
 F(1032)/F(1038)=2-\theta \times{(2\pi-\theta)^4}/1500,
 \end{equation}
\replaced{which is illustrated by the dotted line in Fig.~\ref{fig:denf}.}{which is schematically illustrated in the middle panel and represented by the dotted line in the right panel of Fig.~\ref{fig:denf}.}

Adopting the asymmetric \ion{O}{6} accretion \replaced{flow}{disk} model supplemented by the locally varying $F(1032)/F(1038)$, Monte Carlo simulations were carried out under various parameters, a mass loss rate ${\dot M}$ ranging from $1\times10^{-6}~{\rm M_{\odot}~yr^{-1}}$ to $1\times10^{-5}~{\rm M_{\odot}~yr^{-1}}$ with a step width of $1\times10^{-6}~{\rm M_{\odot}~yr^{-1}}$ and terminal velocities of the Mira wind $v_{\infty}$ with $10~{\rm km~s^{-1}}$, $15~{\rm km~s^{-1}}$, $20~{\rm km~s^{-1}}$ and $25~{\rm km~s^{-1}}$. These parameter sets give the ionization parameter X from 0.2 to 80.

\begin{figure*}
\centering
\includegraphics[scale=0.7]{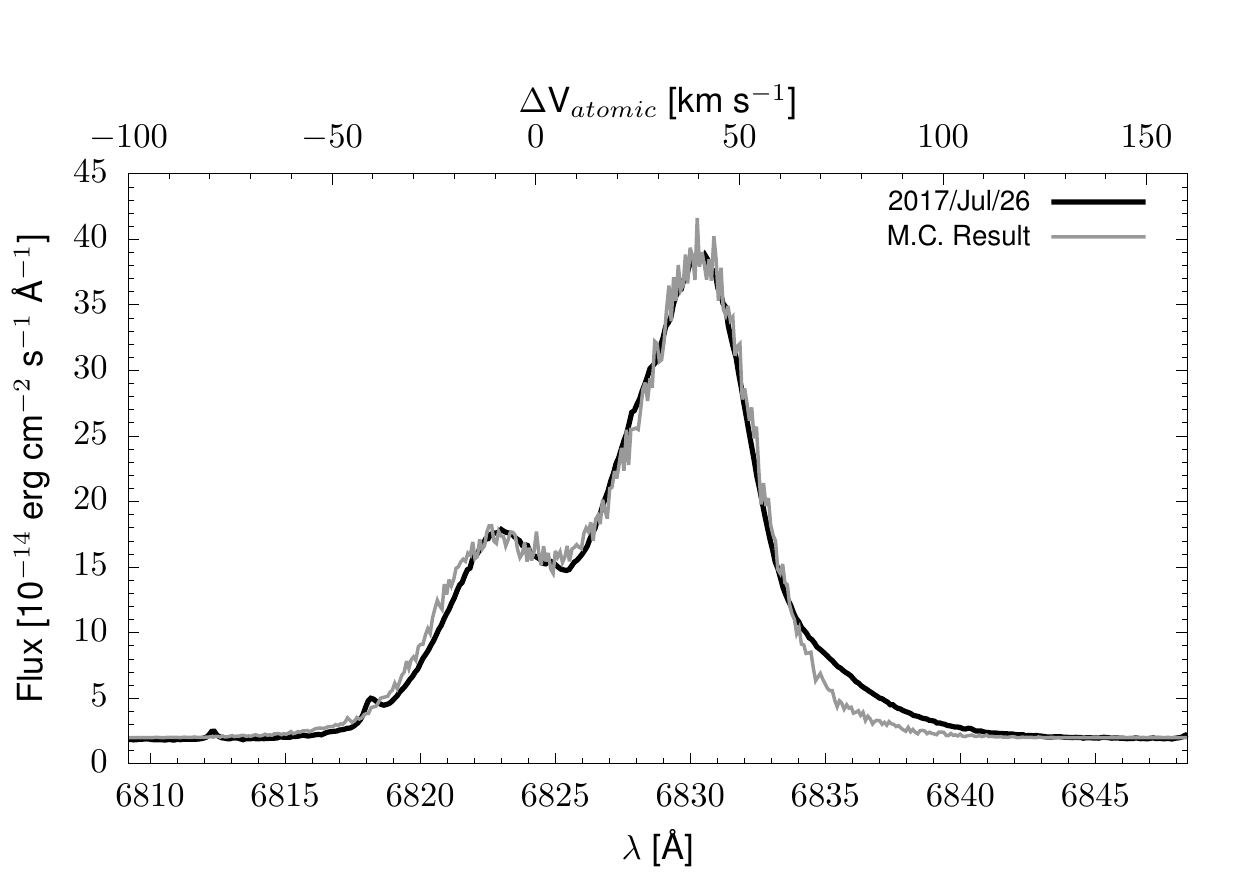}
\includegraphics[scale=0.7]{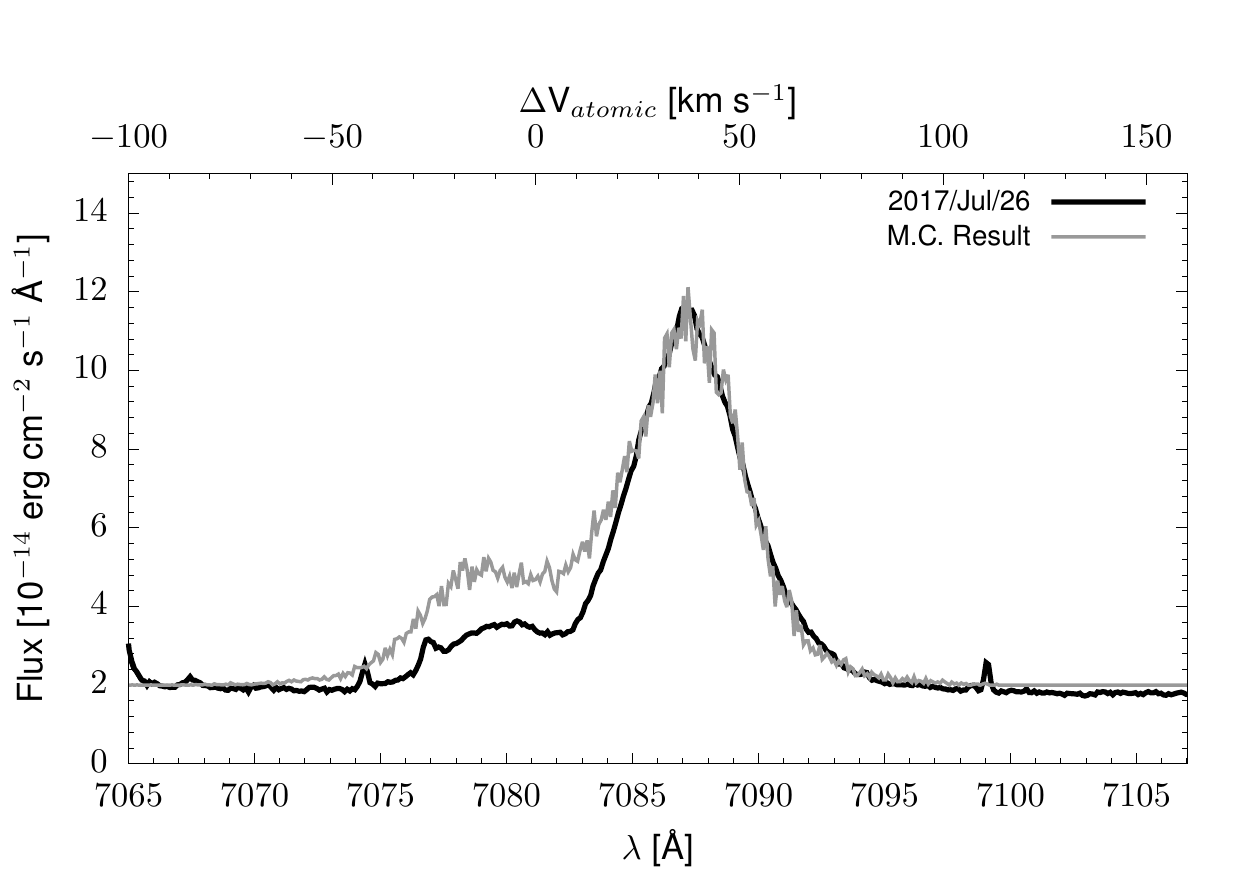}
\caption{Monte Carlo result of Raman-scattered \ion{O}{6} $\lambda\lambda$ 1032 and 1038 features at 6825~\AA\ (left) and 7082~\AA\ (right). The best-fit result is shown in light grey while the observation is shown in black. 
}
\label{fig:sim_ovi}
\end{figure*}

We find that the double-peak structure of Raman \ion{O}{6} features in RR~Tel is diluted if the stellar wind velocity exceeds $v_{\infty}>20~{\rm km~s^{-1}}$ or the ionization parameter $X<1.25$. With the maximum velocity of $v_{\infty}=20~{\rm km~s^{-1}}$, we, therefore, constraint ${\dot M}<8\times10^{-6}~{\rm M_{\odot}~yr^{-1}}$, which gives $X~\sim~1.25$. The corresponding ionization front is presented in Fig.~\ref{fig:model}. The best-fit result is obtained with $v_{\infty}~\sim~20~{\rm km~s^{-1}}$ and ${\dot M}\sim2\times10^{-6}~{\rm M_{\odot}~yr^{-1}}$. In Fig.~\ref{fig:sim_ovi}, we present our best-fit profiles for the Raman \ion{O}{6} features. The black lines show the MIKE data, while the light grey lines represent the results of our simulations. 

Raman scattering efficiency, $\eta$, is defined as the number ratio $N(FUV)/N(Raman)$ of the incident and Raman-scattered photons. The conversion rate for Raman-scattered \ion{O}{6} features was derived from direct measurement of the \ion{O}{6} far-UV emissions in far-UV spectra and \ion{O}{6} Raman features in optical spectra \citep{sc99,birriel00}. Those authors suggested that $N(6825)/N(1032)  \sim 6 \%$ and $N(7082)/N(1038) \sim 2 \%$, respectively. From our Monte Carlo simulations, we obtained $10.5\%$ for the $\lambda1032 \rightarrow \lambda 6825$ conversion and $4.3\%$ for $\lambda1038 \rightarrow \lambda7082$ conversion, respectively.

\begin{table*}
\centering
\caption{Observed far-UV \ion{O}{6} $\lambda\lambda$ 1032, 1038 line fluxes ($F_{obs}$) and expected values ($N_{exp}$, $F_{exp}$) from the best-fit Monte Carlo result. \label{tb:FUV}}
\vskip5pt
 \begin{tabular}{cccccc}
 \hline \hline
Line & $\lambda$ & $\eta$ & $N_{exp}$ & $F_{exp}$ & $F_{obs}$\\
 &(\AA) && ($\rm s^{-1}$) &($\rm erg\ cm^{-2}\ s^{-1}$) &($\rm erg\ cm^{-2}\ s^{-1}$)\\
\hline
 \ion{O}{6} 1032  & 1031.928 & $0.105$ & $5.3 \times10^{46}$ & $1.3\times10^{-9}$ & $2.01\times10^{-10}$\\
 \ion{O}{6} 1038 & 1037.618 & $0.043$ & $3.1 \times 10^{46}$ & $7.3\times10^{-10}$ & $1.18\times10^{-10}$\\
\hline \hline
 \end{tabular}
 \label{tb:ovi}
\end{table*}

For our best-fit results, we obtain the total line flux $F(6825)=3.0\times10^{-12}~\rm erg\ cm^{-2}\ s^{-1}$ and $F(7082)= 0.7\times10^{-12}~\rm erg\ cm^{-2}\ s^{-1}$, respectively. Adopting a distance of $2.6~{\rm kpc}$ \citep{gon13, wh88}, the corresponding number luminosities of Raman-scattered photons emitted per unit time are $N(6825)=5.6~\times~10^{45}~\rm~s^{-1}$ and $N(7082)=1.3~\times~10^{45}~\rm~s^{-1}$. Taking the conversion efficiency for \ion{O}{6}, we deduce the far-UV number flux densities $N(1032)=5.3\times10^{46}~\rm s^{-1}$ and $N(1038)=3.1 \times 10^{46}~\rm s^{-1}$. The corresponding line fluxes are obtained $F(1032)=1.3\times10^{-9}~\rm erg\ cm^{-2}\ s^{-1}$ and $F(1038)=7.3\times10^{-10}~\rm erg\ cm^{-2}\ s^{-1}$, yielding the flux ratio $F(1032)/F(1038)$ of 1.8.
This result does not deviate much from the observed value of 1.7, as shown in Section~\ref{sec:fuse}. In Table~\ref{tb:ovi}, we summarize the intrinsic far-UV \ion{O}{6} parameters including, center wavelength, Raman scattering efficiency $\eta$, calculated number flux density $N_{exp}$, expected line flux $F_{exp}$, and the observed flux $F_{obs}$ in {\it FUSE} data.

\section{\ion{O}{6} $\lambda\lambda$ 3811 and 3834 Doublet} \label{sec:ovi381134}

\begin{figure}
\centering
\includegraphics[scale=0.5]{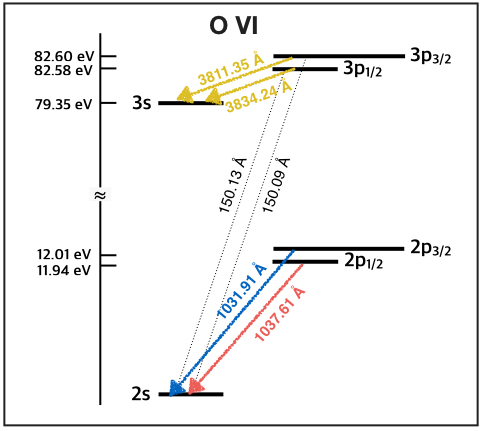}
\caption{Partial Grotrian diagram for \ion{O}{6}. The optical \ion{O}{6}~$\lambda\lambda$ 3811 and 3834 lines arise from the transitions $1s^{2}~3p~(^{2}P^{0}_{3/2,1/2})~\rightarrow~1s^{2}~3s~(^{2}S^{e}_{1/2})$, while the UV resonance doublet \ion{O}{6}~$\lambda\lambda$ 1032 and 1038 correspond to the transitions  $1s^{2}~2s~(^{2}P^{0}_{3/2,1/2})~\rightarrow~1s^{2}~2s~(^{2}S^{e}_{1/2})$, respectively.
}
\label{fig:ovi_gro}
\end{figure}

The \ion{O}{6} $\lambda\lambda$ 3811 and 3834 lines arise from transitions between the level $1s^{2}~3p~(^{2}P^{0}_{3/2,1/2})$ and $1s^{2}~3s~(^{2}S^{0}_{1/2})$, respectively. We present a partial Grotrian diagram for \ion{O}{6} in Fig.~\ref{fig:ovi_gro}. \ion{O}{6} $\lambda$ 3811 appears in the optical line list of RR~Tel by \cite{mc97}, who obtained a high-resolution spectrum in a wavelength range from 3430~\AA\ to 9320~\AA\ with a spectral resolution of R$\sim 25000$. In their spectrum, they found an emission line feature at 3811.35~\AA\ with $\Delta v~\sim~72{\rm~km~s^{-1}}$ and suggested its identification as a blend of \ion{O}{3}~3810.96 and \ion{O}{6}~3811.36. \cite{craw99} extended the line list using data with a higher spectral resolution of R$\sim 50000$. They proposed that the \ion{O}{6}~3834 line could contribute to the emission at 3833.96~\AA, which is detected with $\Delta v~\sim~74{\rm~km~s^{-1}}$ and blended with \ion{He}{2}~3833.80 and \ion{He}{1}~3833.57. \cite{young12} performed an extensive study of \ion{O}{6} recombination lines in RR Tel, from which 19 lines, including \ion{O}{6} 3811 and 3834, were identified. They suggested that a highly ionized region, producing the lines by recombination onto \ion{O}{7}, is situated in the inner structure nearer to the hot WD than the \ion{O}{6} emitting region. 

\begin{table*}
\centering
\caption{Fitting parameters (atomic wavelength, observed wavelength, FWHM, peak value and flux) for \ion{N}{5}, \ion{C}{4}, and \ion{O}{6} doublets.  \label{tb:nvciv}}
\vskip5pt
\begin{tabular}{ccccccc}
\hline\hline
  Line        & $\lambda_{atomic}$ & $\lambda_{obs}$ & FWHM & FWHM & Peak & Flux     \\
          &      (\AA)             &     (\AA)                &   (\AA)    &  (${\rm km~s^{-1}}$)     & (${\rm~erg~cm^{-2}~s^{-1}~{\AA}^{-1}}$) &    ($\rm erg\ cm^{-2}\ s^{-1}$)      \\
\hline
\ion{N}{5} 4603  & 4603.33           & 4602.85             & 0.74  & 48 & $2.5\times 10^{-14}$ & $2.0\times10^{-14}$ \\
\ion{N}{5} 4620  & 4619.78           & 4619.14             & 0.78  & 50 & $8.0\times 10^{-15}$ & $ 6.6\times10^{-15}$ \\
\hline
\ion{C}{4} 5801 & 5801.33           & 5800.22             & 0.83  & 43 & $4.6\times 10^{-14}$ & $4.4\times10^{-14}$ \\
\ion{C}{4} 5812 & 5811.98           & 5810.90             & 0.80  & 42 & $2.4\times 10^{-14}$  & $2.0\times10^{-14}$ \\
\hline
\ion{O}{6} 3811 & 3811.35 & 3810.74 & 0.57 & 44 &  $2.5 \times 10^{-13}$ &  $1.5 \times 10^{-13}$ \\
\ion{O}{6} 3834 & 3834.24 & 3833.67 & 0.57 & 44 &  $1.5 \times 10^{-14}$ & $ 8.9 \times 10^{-14}$ \\
\hline \hline
\end{tabular}
\end{table*}

\begin{figure*}
\centering
\includegraphics[scale=0.3]{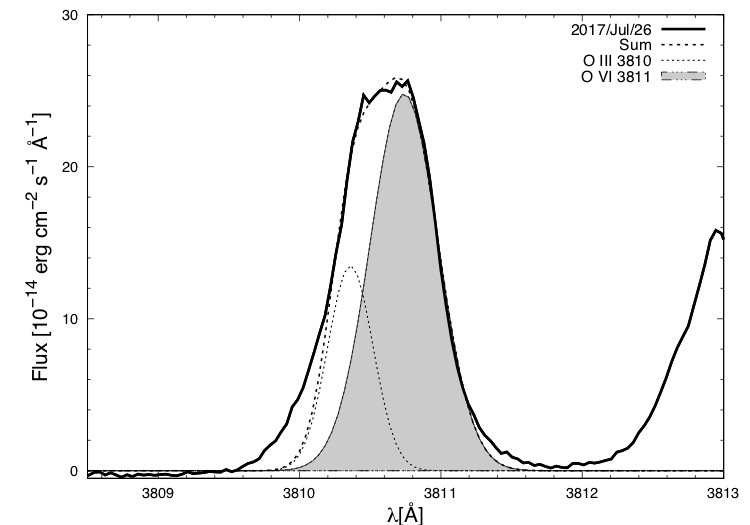}
\includegraphics[scale=0.3]{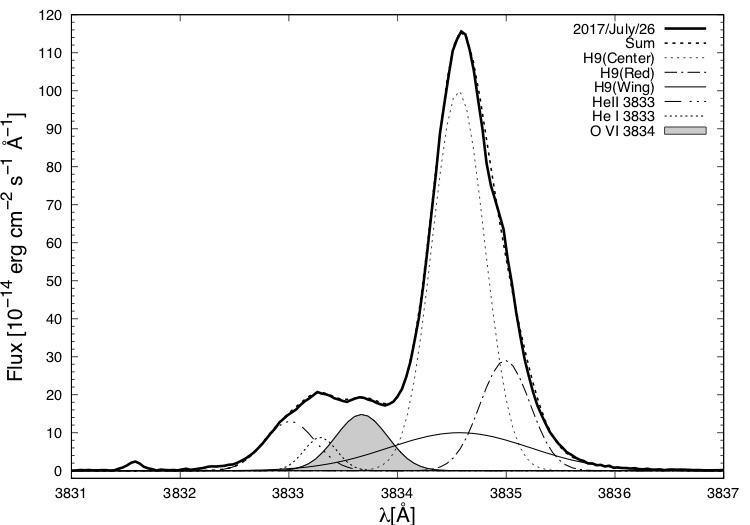}
\caption{\ion{O}{6} $\lambda\lambda$ 3811(left) and 3834(right) doublet in MIKE spectrum. The observed data is shown in thick solid line, and narrow lines represent fitting profiles. Isolated \ion{O}{6} profiles are shown in the grey shades.
}
\label{ovi381134}
\end{figure*}

In this work, we perform a detailed profile decomposition of the emission features at 3811~\AA\ and 3834~\AA\ with our MIKE data and confirm the identification of \ion{O}{6}~$\lambda\lambda$~3811, 3834 doublet. The first step to isolate the \ion{O}{6} doublet from the blended profiles is to infer the profile width of the \ion{O}{6} doublet. It is expected that the profile width of the \ion{O}{6} doublet is comparable to that of \ion{N}{5} $\lambda\lambda$ 4603, 4619 and  \ion{C}{4} $\lambda\lambda$ 5801, 5812 doublets, which have similarly high ionization potential and isoelectronic configurations with \ion{O}{6}. With a single Gaussian function, the \ion{N}{5} doublet lines are located at 4602.85~\AA\ and 4619.14~\AA\ with FWHM  $=48{\rm~km~s^{-1}}$ and $50{\rm~km~s^{-1}}$, respectively. The \ion{C}{4} doublets are found at 5800.22~\AA\ and 5810.90~\AA\ with FWHM  $=43{\rm~km~s^{-1}}$ and $42{\rm~km~s^{-1}}$, respectively. Our fitting parameters with a single Gaussian function for the \ion{N}{5} and \ion{C}{4} doublets are summarized in Table~\ref{tb:nvciv}.

Our MIKE data shows that the 3811 emission feature has a single-peak profile with FWHM $\sim 62~{\rm km~s^{-1}}$. To decompose it into \ion{O}{6}~3811 and \ion{O}{3}~3810, the profile width of \ion{O}{3}~3810 is derived from the adjacent \ion{O}{3} emission line at 3791~\AA, whose profile width is measured to be \replaced{$\sim 24~{\rm km~s^{-1}}$}{$\sim 29~{\rm km~s^{-1}}$}. Within the expected FWHM of $44~{\rm~km~s^{-1}}$ for \ion{O}{6}~3811 and $29~{\rm~km~s^{-1}}$ for \ion{O}{3}~3810, we find that \ion{O}{6}~3811 is centered at 3810.74~\AA\ with a line flux $F(3811) = 1.5\times10^{-13}~{\rm{erg~cm^{-2}~s^{-1}}} $. The fitting result is shown in the left panel of Fig.~\ref{ovi381134}.

As shown in the right panel of Fig.~\ref{ovi381134}, the blended emission feature at 3834~\AA\ has a complicated profile consisting of strong Balmer H9~$\lambda$ 3835, two peaks and a blue shoulder, implying that the 3834 blend comprises at least four emission lines. For decomposition, we first analyze Balmer H9~3835 and \ion{He}{2}~3833 lines by taking H8~3888 \& \ion{He}{2}~3887, and H7~3969 \& \ion{He}{2}~3967 as reference lines. The Balmer line profiles are well fitted using 3 emission components: a central main part, broad wings and a red bump. The Balmer line fluxes are measured to be $F(H7)=2.5\times10^{-12}\rm\ erg~cm^{-2}~s^{-1}$, $F(H8)=1.9\times10^{-12}\rm\ erg~cm^{-2}~s^{-1}$ and $F(H9)=9.3\times10^{-13}\rm\ erg~cm^{-2}~s^{-1}$. The flux ratio $F(H7):F(H8):F(H9)=1:0.78:0.37$ deviates from the case B recombination, which yields a declining trend $1:0.66:0.46$.

\begin{figure*}
\centering
\includegraphics[scale=0.3]{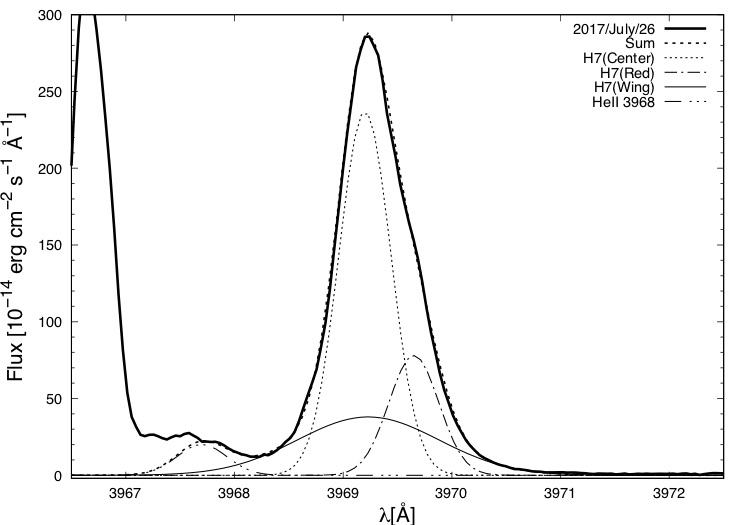}
\includegraphics[scale=0.3]{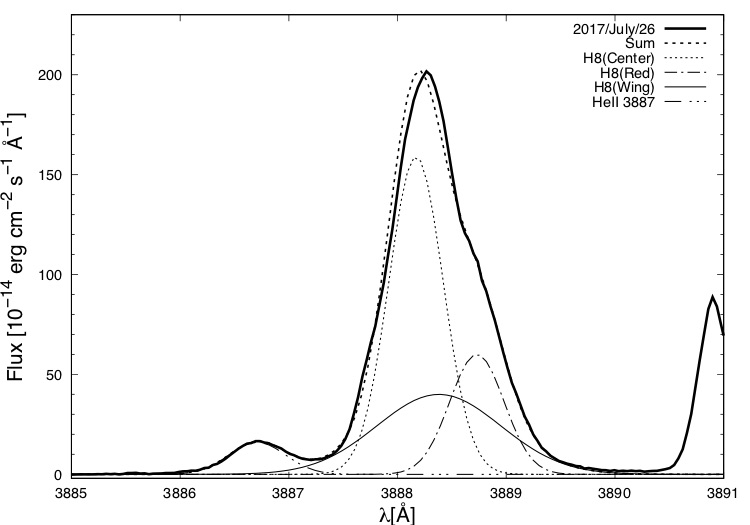}
\caption{Multi-Gaussian fitting of H7~3969 \& \ion{He}{2}~3967 (left) and H8~3888 \& \ion{He}{2}~3887 (right). The MIKE data is shown in thick solid line, and narrow lines represent fitting profiles. 
}
\label{fig:Hprofiles}
\end{figure*}

\ion{He}{2} 3887 and 3967 lines are fitted with a single Gaussian function with FWHM $=0.54$~\AA. Considering the flux ratio of Balmer lines and \ion{He}{2} emissions, the blue shoulder located at around 3833~\AA\ is plausibly identified with He II emission so that we use the term `\ion{He}{2} 3833'. The multi-Gaussian fitting results for the Balmer and \ion{He}{2} emission lines are presented in Fig.~\ref{fig:Hprofiles} and Table~\ref{tb:Hprofiles}.

\begin{table*}
\centering      
\caption{Fitting parameters (center wavelength, FWHM, peak value and flux) for \ion{H}{1}, \ion{He}{2} and \ion{He}{1} emissions. \label{tb:Hprofiles}}
\vskip5pt
 \begin{tabular}{cccccc}
 \hline \hline
    Line                 &                      & Wavelength           & FWHM                & Peak         & Flux                                                            \\
&  & (\AA) & (\AA) & (${\rm~erg~cm^{-2}~s^{-1}~{\AA}^{-1}}$) &    ($\rm erg\ cm^{-2}\ s^{-1}$) \\
\hline
\multirow{3}{*}{H7~3969}  & Center               & 3869.20              & 0.56                 & $2.4\times 10^{-12}$       & $1.4\times 10^{-12}$                  \\
                     & Wing                 & 3969.23              & 1.53                 & $3.8\times 10^{-13}$   & $6.2\times 10^{-13}$    \\
                     & Red                  & 3969.65              & 0.54                 & $7.8\times 10^{-13}$     & $4.5\times 10^{-13}$                                             \\
\ion{He}{2}~3968           &                      & 3967.70              & 0.54                 & $2.0\times 10^{-13}$       & $1.2\times 10^{-14}$                                               \\
\hline
\multirow{3}{*}{H8~3888}  & Center               & 3888.17              & 0.59                 & $1.6\times 10^{-12}$          & $9.9\times 10^{-13}$                                               \\
                     & Wing                 & 3888.38              & 1.34                 & $4.0\times 10^{-13}$           & $5.7\times 10^{-13}$                                   \\
                     & Red                  & 3888.73              & 0.59                 & $6.0\times 10^{-13}$        & $3.8\times 10^{-13}$                                     \\
\ion{He}{2}~3887           &                      & 3886.71              & 0.61                 & $1.6\times 10^{-13}$             & $1.0\times 10^{-13}$                                            \\
\hline
\multirow{3}{*}{H9~3835}  & Center               & 3834.56              & 0.56                 & $1.0\times 10^{-12}$           & $6.0 \times 10^{-13}$                                                 \\
                     & Wing                 & 3834.57              & 1.53                & $1.0\times 10^{-13}$              & $1.6\times 10^{-13}$                                     \\
                     & Red                  & 3834.99              & 0.54                 & $2.9\times 10^{-13}$                 & $1.7\times 10^{-13}$                                   \\
\ion{He}{2}~3833           &                      & 3833.01              & 0.54                 &  $1.3\times 10^{-13}$    & $7.5\times 10^{-14}$   \\
\ion{He}{1}~3833           &                      & 3833.29              & 0.35                 &  $8.8\times 10^{-14}$     & $3.3\times 10^{-14}$  \\
\hline \hline
\end{tabular}
\end{table*}

We find one narrow peak with a small flux at 3833.29~\AA. Using a single Gaussian, the FWHM is measured to be $20{\rm~km~s^{-1}}$, which is smaller than the \ion{O}{6}~3811 line ($44~{\rm~km~s^{-1}}$). In this regard, we suggest that the narrow peak at 3833.29~\AA\ is mainly contributed by \ion{He}{1}~3833 line. 

Taking these three emission lines into account, we finally isolate \ion{O}{6}~3834 with a Gaussian function having the same FWHM ($44~{\rm~km~s^{-1}}$) as that of \ion{O}{6}~3811. We locate the line center at 3833.67~\AA\ and measure a flux of $F(3834) = 8.9\times10^{-14}\rm\ erg~cm^{-2}~s^{-1}$, resulting in $F(3811)/F(3834)\sim 1.7$.

\section{Discussion} \label{sec:dis}

The results obtained from our simulations are based on simple kinematics with Keplerian \replaced{accretion flow}{disk} with an asymmetric matter distribution. \added{\cite{heo16} proposed that the density asymmetry of the accretion disk is responsible for the profile disparity in the Raman \ion{O}{6} features.} The density distribution yielding the best-fit profiles is depicted in Fig.~\ref{fig:denf}, where we note that the highest density on the entering side is 3.5 times that of the most sparse region on the opposite side. In reality, the high contrast of density profile can be made by the elliptical shape of the accretion disk or clumpy, irregular structures in the flow, which should be general features because of the shock front, accretion wake, or the unstable behavior in the \replaced{mass-loss or mass-accretion rates}{wind accretion process}. A number of numerical works of wind accretion in symbiotic stars produced a strong spiral shock and eccentric stream flow, supporting the local density enhancement \citep[e.g.,][]{vkm09,vkm17}. 
\added{However, the 3D spatial information, including its elongation, scale height, and radial motion, etc., is mixed into the Raman \ion{O}{6} profile. To disentangle the properties and geometry of the accretion flow, further modeling combining 3D high-resolution SPH simulations for wind accretion is needed.}

\added{In Fig.~\ref{fig:sim_ovi}, our best fit Monte Carlo profile is insufficient to reproduce the red wing part of the Raman 6825 feature at $\Delta V~\sim~70~{\rm km~s^{-1}}$ only with our accretion \replaced{flow}{disk} model. We point out that the red wing part can be explained with an additional component moving away from the binary orbital plane \citep[e.g.,][]{heo16,lee19}.}

\added{We also note that the blue peak of the Raman 7082  is much below the theoretical range. Taking into account the ratio of Raman scattering cross-section for the \ion{O}{6} 1032 and 1038 photons $\sigma_{Ram}(1032)/ \sigma_{Ram}(1038) = 3.04$ and their oscillation strength $\tau(1032)/\tau(1038)=2$, the acceptable range of the $F(6825)/F(7082)$ should be $<6.08$. However, the MIKE data show that the flux ratio of the blue peaks $f_{blue}(6825)/f_{blue}(7082)$ is 9.7 in 2017. One possible explanation for the relative suppression of the Raman feature was suggested by \cite{sc99}, who proposed the diminution is due to the presence of molecular TiO transitions in the atmosphere and wind of the Mira component.
As discussed in Sec.~\ref{sec:mike}, however, the Mira of RR~Tel is not visible in the optical wavelength. Hence its absorption lines can not be the reason to explain the suppression of the blue portion of the Raman 7082. Further investigation is needed to interpret the discrepancy between the theory and the observation.}

In comparison with the FEROS spectrum taken in 2003, we note that the profiles of Raman \ion{O}{6} features have changed, their red peaks in the MIKE data being stronger. \deleted{According to \cite{heo16}, the density asymmetry of the accretion \replaced{flow}{disk} results in the profile disparity of the Raman \ion{O}{6} features. In this scenario,} The variation of the Raman \ion{O}{6} profiles implies that the density distribution of the accretion \replaced{flow}{disk} has been more asymmetric in the last two decades. The increasing asymmetry of the matter distribution could be related to the dust obscuration episode that occurred between 1996 and 2000, which is associated with the enhancement of the mass-loss rate \citep{kot06,jur12}.  

We also found \ion{O}{6} doublet at 3811~\AA\ and 3834~\AA, which are blended with other emission lines. Our profile decomposition leads us to investigate that the \ion{O}{6} $\lambda\lambda$ 3811, 3834 doublet have a single Gaussian profile with a width of $\sim 44~{\rm km~s^{-1}}$, \added{corresponding to a physical size of $\sim~0.5\rm~au$}. It is worth bearing in mind that the emission region of \ion{O}{6}~$\lambda\lambda$ 1032 and 1038 doublet has a representative \replaced{velocity of $35~{\rm km~s^{-1}}$}{scale of $< 0.8~\rm au$}. This result is consistent with the idea that the \ion{O}{6} recombination lines are produced in the inner part closer to the WD than the \ion{O}{6} resonance lines proposed by \cite{young12}. 
\added{Photoionization modeling is used to obtain the ionization structure of oxygen in the nebula. We conduct a simple exercise for a spherical geometry with a uniform density assuming physical parameters of hot WD, $L_{H}\sim5000~{\rm L_{\odot}}$ and $T_{eff}\sim154000~{\rm K}$, adopted in our work. 
The exercise gives a hydrogen volume density n(H)$\sim10^{9.6}~{\rm cm^{-3}}$.
Our estimation appears to be consistent with 3D hydrodynamical simulations of focused wind accretion in symbiotic stars by \citeauthor{vkm17} (\citeyear[][see their Fig.8]{vkm17}). Their model with $a=8~{\rm au}$, ${\dot M}\sim10^{-6}~{\rm M_{\odot}~yr^{-1}}$, and $v_{wind}\sim 20~{\rm km~s^{-1}}$ produced the accretion stream around WD exceeding the local density $>10^{-16}~{\rm g~cm^{-3}}$. If we adopt the solar abundance, this value is translated into the hydrogen volume density of n(H) $> 10^9 ~{\rm cm^{-3}}$. It should be noted that a direct comparison between our work and theirs is not feasible considering the different parameter sets adopted in this work. }

The identification of \ion{O}{6} $\lambda \lambda$ 3811 and 3834 doublet strongly supports the presence of the EUV doublet at 150.09~\AA\ and 150.13~\AA, originating from $3p \rightarrow 2s$ transitions, as shown in Fig.~\ref{fig:ovi_gro}. Because of a small separation of $\Delta V\sim~80~{\rm km~s^{-1}}$ between two wavelengths, we expect a blended emission at 150.1~\AA\ with a spectrograph having a spectral resolution $R<10000$. \ion{O}{7} is one of the highest observed stages of ionization in the RR~Tel spectrum, and thus can provide a reliable diagnostic of high-temperature plasma. With future EUV observations, our analysis for the \ion{O}{6} and \ion{O}{7} groups could reveal important information on the ionization state and nebular physics in RR~Tel. 

\section{Summary}\label{sec:sum}
RR~Tel is a symbiotic nova, consisting of a Mira variable and a hot WD. It is known as one of the objects having broad emission features at 6825~\AA\ and 7082~\AA, which originate from Raman scattering of \ion{O}{6} 1032 and 1038 by neutral hydrogen atoms. In this work, we present optical high-resolution spectra of RR~Tel obtained with MIKE and FEROS and the far-UV {\it FUSE} spectrum. 
\replaced{With the WD mass of $M_{WD}=0.65~{\rm M_{\odot}}$, the representative scale of the \ion{O}{6} emission region is estimated to be $< 0.8 \rm~au$, which gives a corresponding velocity of $35~{\rm km~s^{-1}}$. The Raman best-fit profile is obtained with a mass loss rate of the Mira ${\dot M}\sim2\times10^{-6}~{\rm M_{\odot}~yr^{-1}}$ and a wind terminal velocity $v_{\infty}\sim 20~{\rm km~s^{-1}}$. Our profile analysis implies the possible presence of the bipolar outflow component receding $\Delta~V~\sim~70~{\rm km\ s^{-1}}$ in RR~Tel. We have performed a profile analysis of the Raman \ion{O}{6} features by assuming that the \ion{O}{6} emission traces the Keplerian flow around the WD.}
{We have performed Monte Carlo simulations for the profiles of the Raman \ion{O}{6} features, assuming a Keplerian disk. It allows us to map the \ion{O}{6} disk and estimates the representative scale of $< 0.8 \rm~au$. 
The best-fit profiles are obtained with a mass loss rate of the Mira ${\dot M}\sim2\times10^{-6}~{\rm M_{\odot}~yr^{-1}}$ and a wind terminal velocity $v_{\infty}\sim 20~{\rm km~s^{-1}}$.
Comparison with FEROS data provides evidence of a change in the density asymmetry of the \ion{O}{6} disk, which can be associated with the mass transfer history of RR~Tel in the last two decades. 
We also identify \ion{O}{6} $\lambda \lambda$ 3811 and 3834 doublet from detailed profile decomposition, from which  the physical scale of the \ion{O}{7} region is estimated to be $\sim0.5~{\rm au}$. 
It is emphasized again that Raman profiles provide essential information regarding the matter distribution and kinematics of the \ion{H}{1} and \ion{O}{6} components leading to the advancement of our knowledge on the stellar wind accretion in symbiotic stars. Combining 3D hydrodynamical modeling might improve our overall understanding of the WRLOF mode and the accretion phenomena in D-type symbiotic systems.
}

\begin{acknowledgments}
We would like to thank the anonymous referee for constructive comments that greatly improved this paper.
We are also grateful to Young-Min Lee for valuable discussions.
This research was supported by the Korea Astronomy and Space Science Institute under the R\&D program (Project \#2018-1-860-00) supervised by the Ministry of Science, ICT, and Future Planning. 
J.E.H is supported through a Gemini Science Fellowship by the international Gemini Observatory, a program of NSFs OIR Lab, which is managed by the Association of Universities for Research in Astronomy (AURA) under a cooperative agreement with the National Science Foundation, on behalf of the Gemini partnership of Argentina, Brazil, Canada, Chile, the Republic of Korea, and
the United States of America. J.E.H was also supported from Basic Science Research Program through the National Research Foundation of Korea (NRF) funded by the Ministry of Education (\#2019R1A6A3A03033787).
R.A. acknowledges financial support from DIDULS Regular PR\#1953853 by Universidad de La Serena.
This paper includes data gathered with the 6.5 meter Magellan Telescopes located at Las Campanas Observatory, Chile. 
This work is based on observations collected at the European Southern Observatory under ESO programme ID 60.A-9120(B).
\end{acknowledgments}


\vspace{5mm}
\facilities{Magellan:Clay (MIKE), Max Planck:2.2m (FEROS), {\it FUSE}}


\software{IRAF \citep{tody86,tody93}, CarPy \citep{kelson00,kelson03}}




\bibliography{sample631}{}
\bibliographystyle{aasjournal}



\end{document}